\documentstyle[12pt,graphicx]{article}
\def\v{\begingroup\obeyspaces\u}
\def\u#1{\verb!#1!\endgroup}

\def\HW{{\small Herwig++}}
\def\FHW{{\small HERWIG}}

\def\DY{{\small DOXYGEN}}

\begin{document}
\tolerance=100000
\thispagestyle{empty}
\setcounter{page}{0}
 \begin{flushright}
Cavendish-HEP-06/05\\
CERN-PH-TH/2006-021\\
IFJPAN-IV-2006-2\\
IPPP/06/09\\
KA-TP-02-2006\\
February 2006
\end{flushright}

\begin{center}
{\Large \bf Herwig++ 2.0\boldmath{$\beta$} Release Note}\\[4mm]

{S. Gieseke\\[0.4mm]
\it Institute f\"{u}r Theoretische Physik, Karlsruhe\\[0.4mm]
E-mail: \tt{gieseke@particle.uni-karlsruhe.de}}\\[4mm]

{D.\ Grellscheid\\[0.4mm]
\it IPPP, Department of Physics, Durham University\\[0.4mm]
E-mail: \tt{david.grellscheid@durham.ac.uk}}\\[4mm]

{A.\ Ribon\\[0.4mm]
\it  PH Department, CERN\\[0.4mm]
E-mail: \tt{Alberto.Ribon@cern.ch}}\\[4mm]

{P.\ Richardson\\[0.4mm]
\it IPPP, Department of Physics, Durham University\\[0.4mm]
E-mail: \tt{Peter.Richardson@durham.ac.uk.ac.uk}}\\[4mm]

{M.H.\ Seymour\\[0.4mm]
\it PH Department, CERN\\[0.4mm]
E-mail: \tt{M.Seymour@rl.ac.uk}}\\[4mm]

{P.\ Stephens\\[0.4mm]
\it Institute of Nuclear Physics, Cracow\\[0.4mm]
E-mail: \tt{stephens@hep.phy.cam.ac.uk}}\\[4mm]

{B.R.\ Webber\\[0.4mm]
\it Cavendish Laboratory, University of Cambridge\\[0.4mm]
E-mail: \tt{webber@hep.phy.cam.ac.uk}}\\[4mm]

\end{center}

\vspace*{\fill}

\begin{abstract}{\small\noindent
    A new release of the Monte Carlo program \HW\ (version 2.0$\beta$) is now
    available. The main  new  feature is the extension of the program to 
    include simple hadron-hadron processes including the initial-state parton
    shower.}
\end{abstract}

\vspace*{\fill}
\newpage
\tableofcontents
\setcounter{page}{1}

\section{Introduction}

The last major public version (1.0) of \HW\ was reported in detail
in \cite{Gieseke:2003hm}. In this note we describe the main modifications
and new features included in the latest public version, 2.0$\beta$.

Please refer to \cite{Gieseke:2003hm} and to the present paper if
using version 2.0$\beta$ of the program.

\subsection{Availability}
The new program, together  with other useful files and information,
can be obtained from the following web site:
\small\begin{quote}\tt
            http://hepforge.cedar.ac.uk/herwig/
\end{quote}\normalsize

\section{Hadron-Hadron Collisions}

\begin{figure}[t]
\includegraphics[angle=90,width=0.45\textwidth]{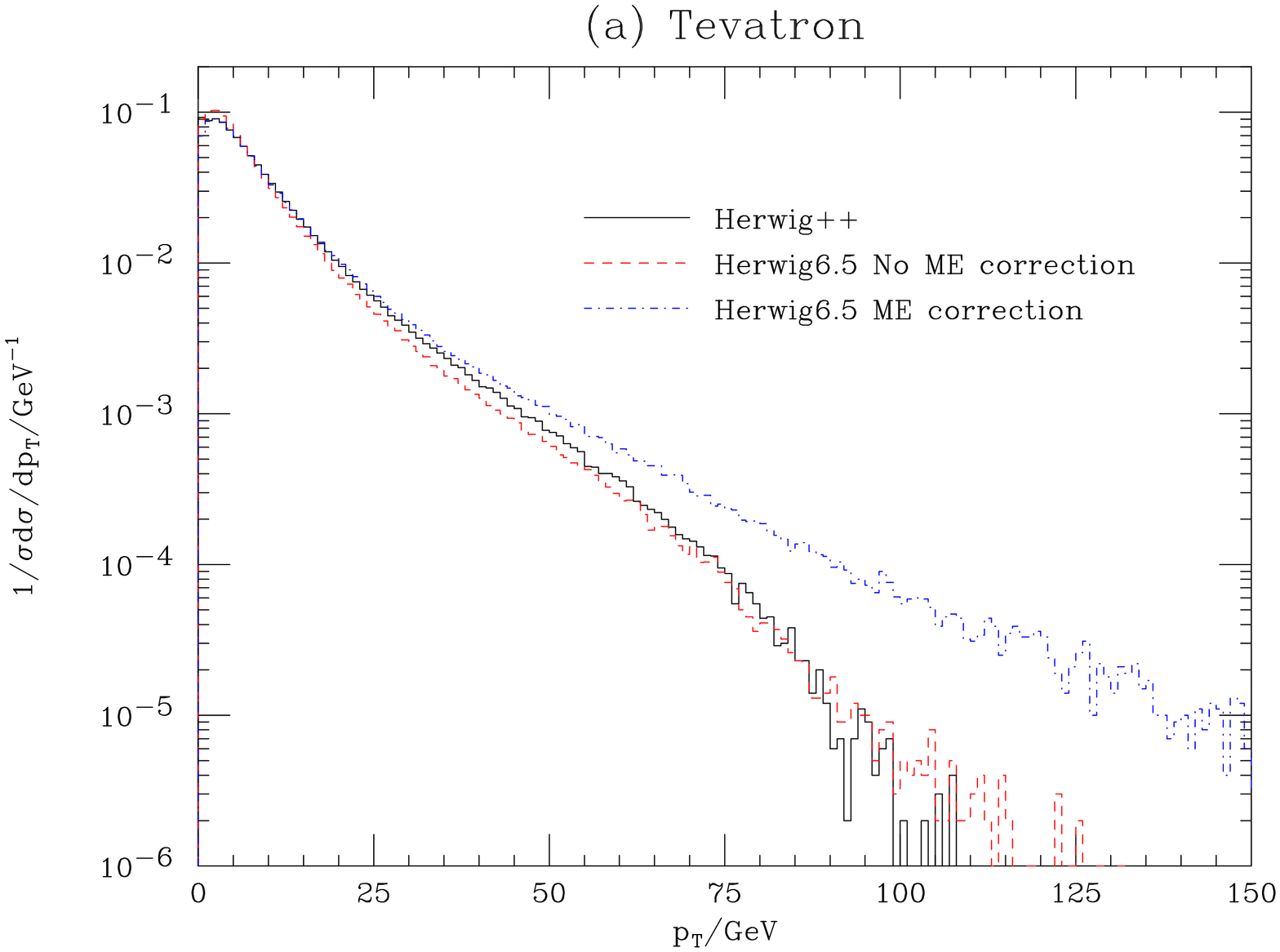}\hfill
\includegraphics[angle=90,width=0.45\textwidth]{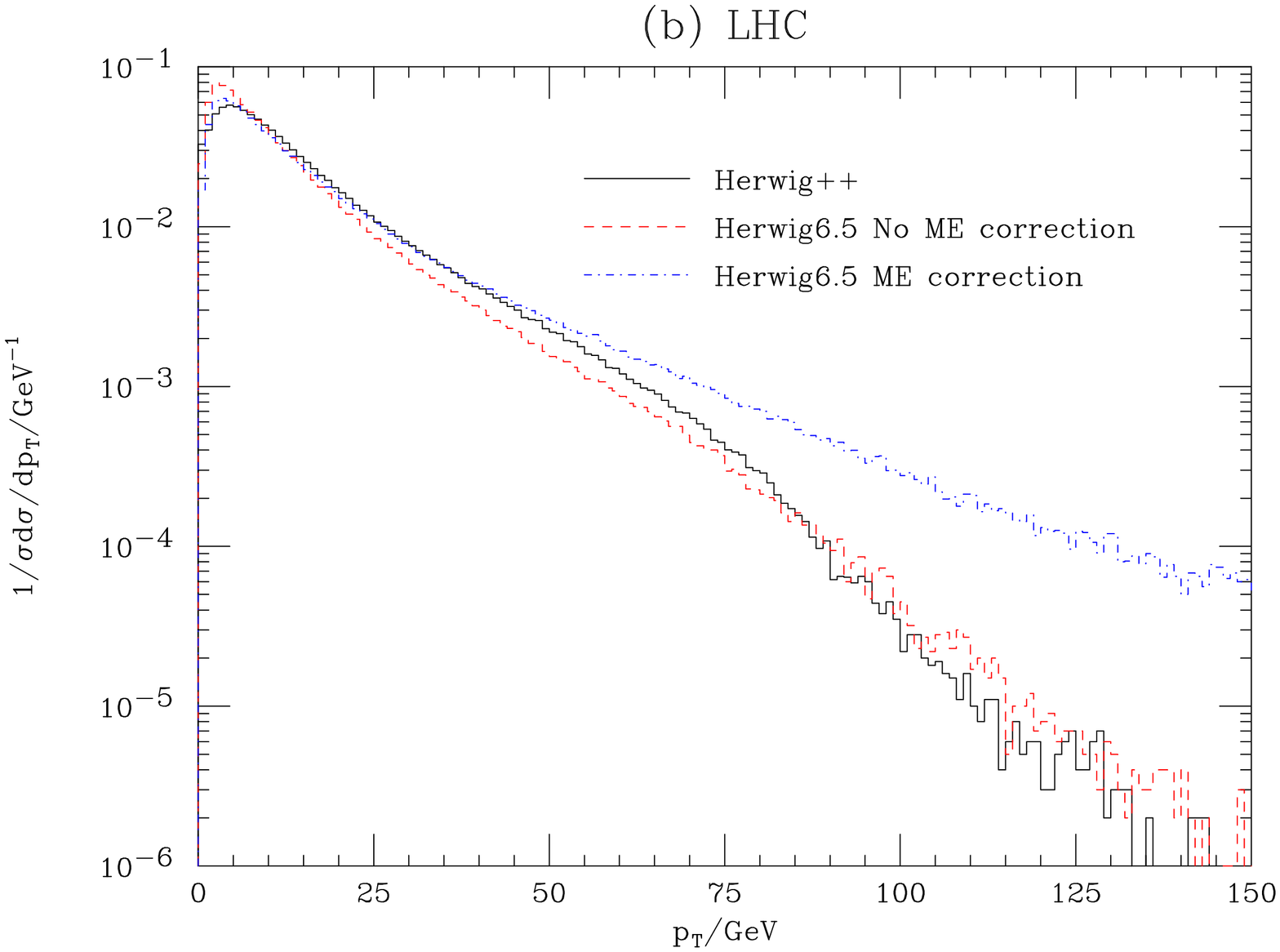}
\caption{The $p_T$ spectrum of $\gamma^*/Z$ bosons produced at (a) the Tevatron and (b) the LHC using \HW2.0$\beta$ compared
         with \FHW6.5 with and without matrix element correction. In both cases the mass of the Drell-Yan pair was required
         to be greater than $20$\,GeV.}
\label{fig:Zpt}
\includegraphics[angle=90,width=0.45\textwidth]{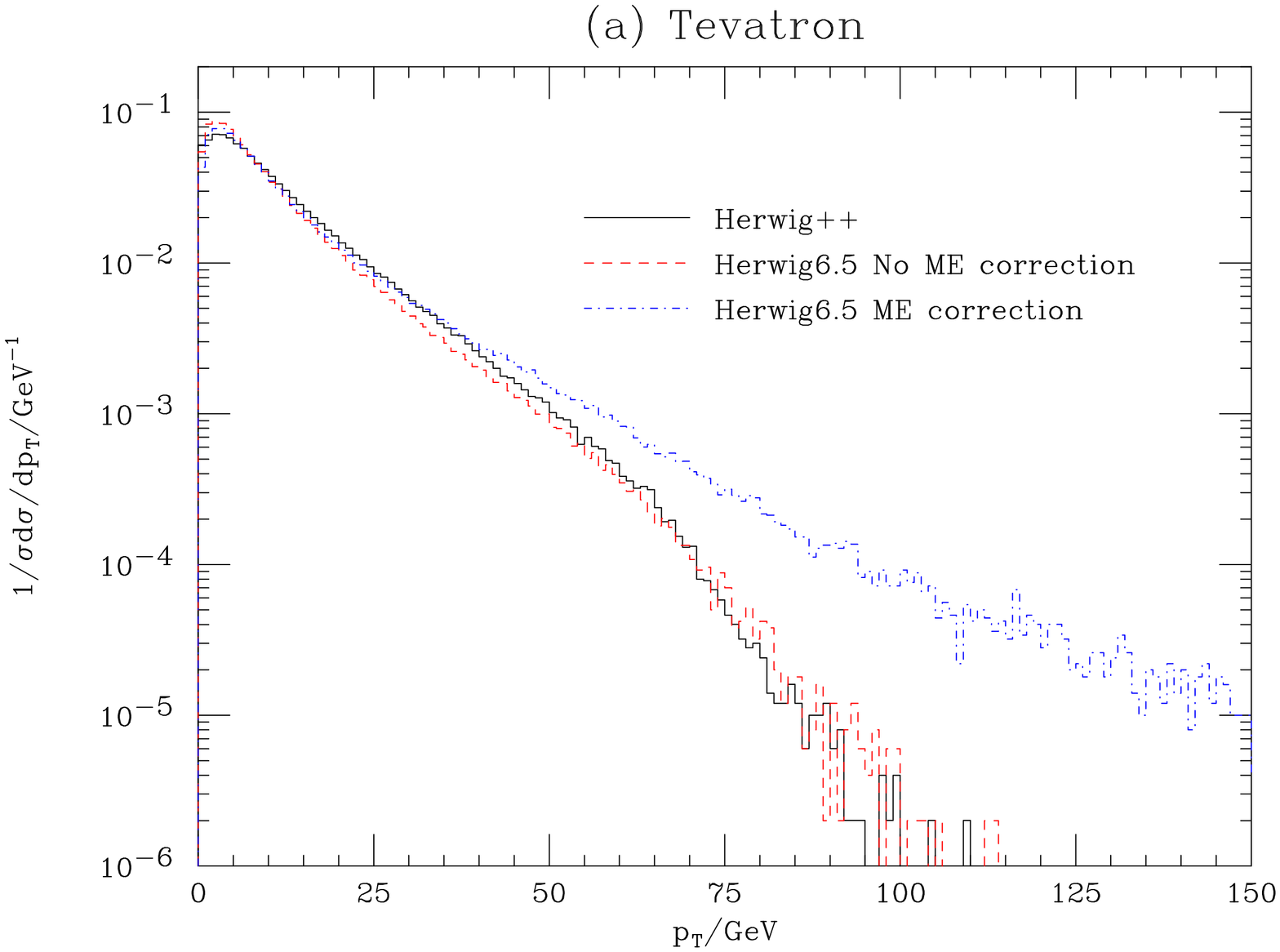}\hfill
\includegraphics[angle=90,width=0.45\textwidth]{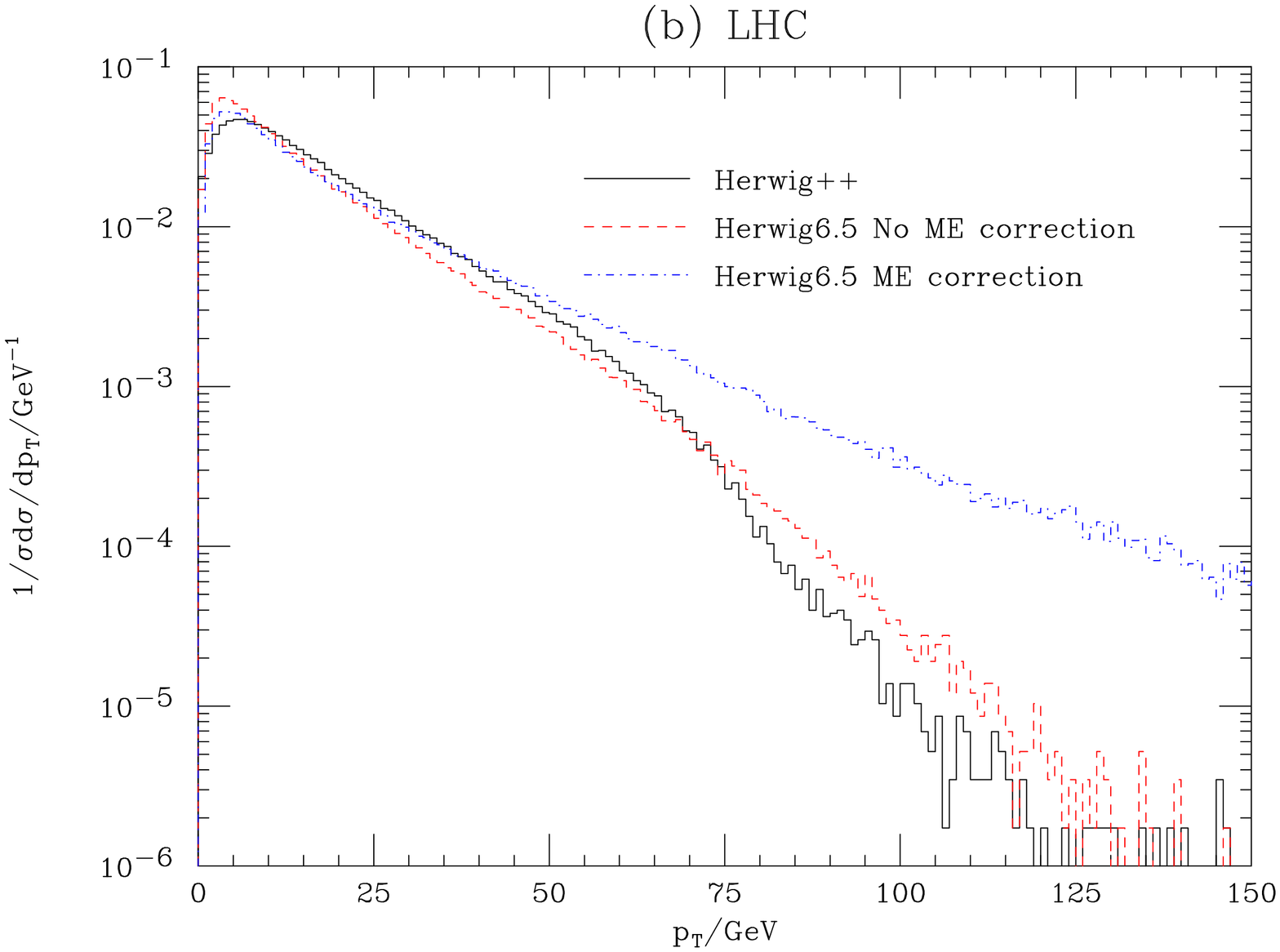}
\caption{The $p_T$ spectrum of $W$ bosons produced at (a) the Tevatron and (b) the LHC using \HW2.0$\beta$ compared
         with \FHW6.5 with and without matrix element correction.}
\label{fig:Wpt}
\end{figure}

  The main new feature of this version is the extension of the original
  ${\rm e}^+{\rm e}^-$ program to hadron-hadron collisions. In the current
  version only simple Drell-Yan processes~(both $W$ and $Z$ production) are supported
  together with the initial-state shower from the incoming partons, using the
  algorithm described in \cite{Gieseke:2003rz}. The
  outgoing partons radiated from the incoming lines are not currently showered, 
  there is no model of the underlying event or matrix element correction
  for Drell-Yan processes. However, this version does produce 
  hadron-hadron events which can be used to test the integration of the program
  into experimental simulations and gives an improved $p_T$ spectrum of the gauge
  bosons with respect to the FORTRAN \FHW\ 6.5 without matrix element corrections.
  The $p_T$ distributions of $Z$ and $W$ bosons, at the Tevatron and LHC, are compared to \FHW6.5
  in Figs\,\ref{fig:Zpt} and \ref{fig:Wpt} respectively.

\section{Other Changes}

\begin{itemize}
\item A number of changes to the decays and hadronization have been made to improve the
      stability of the code. As a result of these changes the default strange quark
      weight \v{PwtSquark} has been returned to its natural value of 1.0, which increases the amount of K mesons
      produced and improves the agreement with LEP data.
\item The documentation has been changed to use \DY\ and is available either via
      the \HW\ web-page or with the code. In addition information on using the program 
      has been added to the \HW\ wiki.
\item The build procedure has been significantly improved and now uses the GNU autotools.
\item A large library of helicity classes has been included to make implementing
      additional matrix elements much simpler.
\item Many models for hadronic decays have been added which are currently not used by
      default but can be switched on if needed (see the web-site for how to do this).
\end{itemize}

\end{document}